# Evaluation of compliance with democratic and technical standards of i-voting in elections to academic senates in Czech higher education

## Tomáš Martínek

*Author ORCID Nr: 0000-0002-4295-0886*
*Department of Information Engineering, Faculty of Economics and Management, Czech University of Life Sciences Prague, Czech Republic, martinekt@pef.czu.cz*

## Michal Malý

*Author ORCID Nr: 0000-0002-4782-2672*
*Institute of Political Studies, Faculty of Social Sciences, Charles University, Czech Republic, michal.maly@fsv.cuni.cz*

*Abstract: The shift towards increased remote work and digital communication, driven by recent global developments, has led to the widespread adoption of i-voting systems, including in academic institutions. This paper critically evaluates the use of i-voting platforms for elections to academic senates at Czech public universities, focusing on the democratic and technical challenges they present. A total of 18 out of 26 Czech public universities have implemented remote electronic voting for these elections. Yet, the systems often lack the necessary transparency, raising significant concerns regarding their adherence to democratic norms, such as election security, voter privacy, and the integrity of the process. Through interviews with system developers and administrators, along with a survey of potential voters, the study underscores the critical need for transparency. Without it, a comprehensive assessment of the technical standards and the overall legitimacy of the i-voting systems remains unattainable, potentially undermining the credibility of the electoral outcomes.*

## 1. Introduction

Universities have been pioneers in digitizing various processes, and this trend has been accelerated by the COVID-19 pandemic, which necessitated greater remote communication via online tools. As part of this digital transformation, universities have integrated remote internet voting (i-voting) for their electoral processes, including the election of academic senate members.

The Magna Charta Universitatum of 1988 already established the fundamental principles of university autonomy, emphasizing that universities should be independent of political and economic power. The Czech Higher Education Act (Czech Republic 2024) further guarantees the democratic rights of university members to elect their representative bodies through direct and secret elections.

The primary goal of this paper is to analyze the use of i-voting systems at Czech public universities in the context of their compliance with democratic norms, such as transparency, security, representation, and participation. The paper applies the evaluation framework proposed by Von Nostitz et al. (2021) to assess how well these systems meet both democratic and technical standards across a sample of Czech universities.

## 2. Literature review

### 2.1. Academic freedoms

Already in 1988, in Bologna, during the celebration of the 900th anniversary of the world's oldest university, university rectors signed the basic principles (Magna Charta Universitatum, 2022) on which the role of universities must be based. Among the signatories was the rector of Charles University in Prague, which at that time was still under the rule of a totalitarian communist regime that would not have considered the given charter text compatible with the state ideology (Jařab, 2008). According to the 1988 Magna Charta, in terms of democracy, a university is an autonomous institution where research and teaching are morally and intellectually independent of any political and economic power. Every university must guarantee its students basic freedoms. The Czech Higher Education Act (Czech Republic, 2024) guarantees the academic right of members of the academic community to elect representative academic bodies. The Academic Senate of a public university is its self-governing representative academic body with at least 11 members, of which at least one-third and no more than half are students. The powers of the Academic Senate include, inter alia, deciding on the establishment, merger, amalgamation, division or abolition of units of the university, approving internal regulations, the budget, and annual reports, and deliberating on a proposal for the appointment of the Rector or a proposal for his/her removal from office.

### 2.2. I-voting at world universities

Voting is a fundamental element of democracy, and it is crucial that elections are not only free but also fair. The secrecy of the ballot is essential as it ensures that each voter can cast their vote without fear of coercion or retribution. To maintain the fairness of elections, it is imperative that they operate on the principle of "one person, one vote" and that they are safeguarded against manipulations such as multiple voting or the destruction of valid ballots (Bishop and Hoeffler 2016). This is particularly important in the context of i-voting platforms, where these principles must be rigorously upheld to ensure the legitimacy and credibility of the tool. In the academic environment, it can be used for secret elections to the Academic Senate or College Councils, for secret ballots of councils or committees, voting for Dean's and Rector's awards, anonymous point evaluation of teaching, within student associations, and elsewhere. Although the effects of voting in universities are not as significant as in national political elections, academic senates still have some significant powers. Therefore, even i-voting in universities should meet at least the basic security requirements (Qadah & Taha, 2007) to ensure the integrity and credibility of the entire election system. I-voting systems are voting systems that allow remote casting of votes over the Internet in an uncontrolled environment using a client device. A broader group are e-voting systems, which include other methods of electronic voting. I-voting systems are used in universities primarily to save time and improve convenience for voters and committee members, eliminate errors in filling out the ballot, speed up the counting process, and increase the accessibility of independent voting (Pawar et al., 2020). Accessible i-voting systems also have the advantage of allowing blind and otherwise disabled people to vote independently (Huarte et al., 2008). In Estonia, i-voting has also been calculated to be the cheapest way to cast a vote in a multi-channel voting system (Krimmer et al., 2018), which also has the lowest environmental impact (Willemson & Krips, 2023). The tradition of remote electronic voting at Czech universities is quite long, with the Academic Senate being elected electronically at the Faculty of Science of Charles University in Prague as early as 2003 (AS PřF UK, 2007).

The implementation of a secret and verifiable Internet-based voting system in a university setting was described in detail by Adida et al. (2009), where the author of the Helios Voting Adida system worked with academics during the election of the Rector of the Belgian Catholic University of Louvain (UCL) in Louvain-la-Neuve in March 2009. The open-source Helios system uses a modified exponential asymmetric encryption of ElGamal (1985) with an additive homomorphic property for vote secrecy, which allows aggregation of votes by multiplying the encrypted votes without having to decrypt them individually. The individual votes are encrypted while still in the voter's device before being sent to the electronic ballot box of the Helios system. Distributed decryption is used to decrypt the overall results, where multiple trustees' keys are needed. Of the 25,000 eligible voters at the University of Leuven, 5,000 registered to vote, with nearly 4,000 voters voting in each of the two rounds.

A security analysis of the use of the Helios Voting system in a university setting was conducted by Junior et al. (2022), who focused on its use in Brazilian universities - mainly the Federal University Agreste de Pernambuco in 2020, as well as other universities including the University of São Paulo, the Federal University of Pelotas, and the Federal University of Technology Paraná.

Other well-documented (Baloglu et al., 2021) open-source i-voting systems used in universities around the world include software developed by academics at the University of Lorraine - Belenios (Cortier et al., 2019), which also complies with the second level security requirements of the French CNIL recommendations (Cortier et al., 2020).

Among the commercial tools, the Invote system from the Spanish company Scytl (Finogina et al., 2024), which is used at the University of Zaragoza, the Autonomous University of Barcelona or the University of Oviedo (Scytl, 2024), is well known and academically evaluated (Marcos del Blanco & Gascó, 2019). Another Spanish i-voting system evaluated (Marcos del Blanco et al., 2020) is Nvotes, which is used at the National University of Distance Education, the second-largest university in Europe by student numbers (nVotes, 2024).

## 2.3. Academic senates

Academic senates play a key role in university governance, providing a platform for faculty involvement in decision-making processes, especially in academic matters. They are responsible for setting academic policies, approving new courses, and overseeing the quality of education (Jones, Shanahan, & Goyan, 2002). The structure of academic senates varies among institutions but typically reflects a bicameral system where the senate handles academic matters while the board of trustees handles administrative and financial matters. This system aims to balance academic and administrative interests, although there is often tension and confusion in roles (Jones, Shanahan, & Goyan, 2002).

A number of studies highlight several challenges facing academic senates, including role ambiguity, limited influence over budgetary matters, and difficulties in adapting to rapid policy changes (Pennock et al., 2015; Birnbaum, 1989). Bradshaw and Fredette (2008) highlight the complexity of academic governance and the need for strategic management. Birnbaum (1989) identifies the latent organizational functions of academic senates in his study "The Latent Organizational Functions of the Academic Senate". He argues that senates, while often not functioning as expected, perform important symbolic and stabilizing roles that contribute to their longevity despite their apparent failures in some areas. A study by Pennock et al. (2015) analyzes changes in the structure and role of academic senates in Canadian universities between 2000 and 2012. The results point to the need for improved orientation programs for senators, better oversight of academic quality, and clarification of roles between the senate, the board, and the administration.

Jones, Shanahan, and Goyan (2002) examine the effects of fiscal restraint and market-based research funding on the governance of Canadian universities, focusing in particular on the role and effectiveness of university senates. The research shows that senates face a disconnect between their intended and actual roles in academic decision-making, highlighting the challenges these bodies face in the context of contemporary political pressures.

Kouba's study focuses on the determinants of student participation in academic senate elections at public universities in the Czech Republic. The research applies the rational choice model to academic elections, which is relatively uncommon, and examines how various factors influence students' decisions to participate in elections for bodies that can have a significant impact on the governance of universities (Kouba, 2017). He finds that smaller academic units tend to have higher turnout in elections, which the author interprets as a result of a stronger sense of community and personal interest in election outcomes. Conversely, larger units often show lower participation, which may be due to anonymity and a lower sense of accountability to the academic community. The competitiveness of elections leads to greater participation, and technological innovations such as Internet voting have a positive but less significant effect on participation. This research offers important insights into how students choose to participate in academic elections and what factors institutions can influence to increase student engagement in college governance (Kouba, 2017).

## 2.4.   Digital Democracy Platforms

Online and internet voting (i-voting) represents a significant advance in modern political participation, allowing for greater involvement of party members and voters in decision-making processes.

Deseriis (2021) highlights that digital participation platforms, including those that enable i-voting, reduce the costs of political participation, thereby enabling greater participation of individuals. Deseriis analyses the different strategies by which designers and administrators of decision-making software such as LiquidFeedback, Consul, and others facilitate voting processes for users. These platforms are designed to improve stakeholder participation, give them access to decision-making processes, and enhance political equality (Deseriis, 2021).

Von Nostitz and Sandri (2021) focus on the digitalization of electoral processes within political parties. They identify several key indicators of successful i-voting, including accessibility and user-friendliness, data security and protection, regulation and platform governance, integration, and communication among members. These indicators are essential for ensuring member confidence in the voting process and preventing manipulation of results (Von Nostitz & Sandri, 2021).

Paolo Gerbaudo (2019) analyses the democratic quality of online participation platforms in the Five Star Movement in Italy and Podemos in Spain. He finds that despite the promises of these digital parties to introduce more member-driven and inclusive democracy, internal democracy in digital parties is highly centralized and plebiscitary in nature. Members have limited influence over decision-making processes and platforms are tools for affirming leadership decisions rather than for genuine deliberation and participation (Gerbaudo, 2019).

In his comparative analysis of the technopolitical cultures and participation platforms of the Pirate Party of Germany and the Five Star Movement, Deseriis (2020) shows how different technopolitical cultures influence the organization of these parties and how the subsequent adoption and use of online participation platforms has led to internal conflicts and disputes within the Pirate Party and increased centralization within the Five Star Movement. This research highlights the importance of cultural context and organizational practices in the implementation of i-voting (Deseriis, 2020).

Overall, online and web-based voting has the potential to transform political participation and decision-making, but its success depends on a number of factors, including software design, regulatory standards, and how they are implemented and used by political actors. Gerbaudo (2021) describes that for digital participatory platforms to be effective, they must be transparent and have clearly defined rules of operation; otherwise, there is a risk of losing legitimacy and voter passivity. Further empirical research is needed to understand the specific impacts of these technologies on political equality and the effectiveness of decision-making processes (Deseriis, 2021; Von Nostitz & Sandri, 2021; Gerbaudo, 2019). Based on these findings, our study will evaluate online voting for academic senate elections using the concepts of Von Nostitz and Sandri (2021).

3.  **Research methods**

Analysis of relevant articles was conducted from the scientific databases Web of Science and Scopus. Based on those, our analysis of the use of remote electronic systems at Czech public universities for academic senate elections is conducted using online interviews with administrators and developers of IS of Czech public universities, publicly available documents, system testing, and available source code.

## 3.1.  **Methodology for evaluation of the systems of Czech public universities**

The semi-structured interviews ask questions about system development and administration, verifiability, confidentiality and security of the election, voter authentication, resistance to coercion, integrity, voter awareness, and how to deal with possible problematic situations where, for example, a voter could vote for both student and staff senate chambers at the same time. Representatives of all 26 Czech public universities were contacted and structured interviews were conducted with representatives of four Czech public universities that use the four most commonly used i-voting systems, namely from DERS Group, IS4U, Faculty of Informatics Masaryk University in Brno (FI MUNI) and Computer Center Palacky University Olomouc (CVT UPOL). The paper discusses in detail a sample of i-voting systems used at more than one public university, which includes 4 i-voting systems evaluated through interviews supplemented by the Elections system from Faculty of Arts Charles University in Prague (FA CUNI).

For a comprehensive evaluation of the i-voting systems of Czech universities, the concept proposed by Von Nostitz et al. (2021) is used, which defines an evaluation framework for democratic norms including participation, representation, competitiveness, responsiveness, and transparency. It further defines an evaluation framework for technical standards including security, verification process, data ownership, and control. For the framework, a descriptive analysis of voting systems in Czech public universities is conducted based on research findings. Specifically, **democratic norms** for academic environments are specified by the author:

**Participation** of voters (members of the academic community) in voting, which is influenced by the degree of inclusiveness and accessibility. All eligible voters should be able to vote through one of the channels, while respecting constitutional principles such as universality or secrecy. In the framework of the participation evaluation, data from the Faculty of Arts at Charles University in Prague is used (Charles University, 2024). This faculty has the highest number of full-time students among the faculties using i-voting in the Czech Republic  (Ministry of Education of the Czech Republic, 2024).

**Representation** in the context of academic senate elections refers to the extent to which election outcomes reflect the diverse composition of the university community, including students, faculty, and staff from different faculties and departments. The i-voting system should ensure equal access for all eligible voters, regardless of their socio-economic status, technical resources, or other barriers, to ensure that the election results truly represent the voices of all groups, including minority or marginalized constituencies within the academic environment.

**Competitiveness** ensures that elections allow fair competition among candidates and that all candidates have equal opportunities to participate. In i-voting systems for academic senates, it is crucial to ensure that candidates from different faculties and academic positions are not given unequal advantages and that the system does not favor certain groups or individuals, promoting a fair electoral process.

**Transparency and accountability** means that the entire e-voting process is open to independent audit and verification. In internet voting systems, transparency can be ensured through the use of open source software or independent audits to verify that the election has not been rigged and that the results accurately reflect the will of the voters. Transparency and accountability should guarantee the legitimacy of the election results. This ensures voter confidence in the system and acceptance of Internet voting as a democratic and legitimate voting method.

**Responsiveness** refers to how quickly and efficiently issues are addressed and how voters' preferences are represented by elected representatives. In the context of i-voting for academic senates, it also means that universities should have mechanisms in place to quickly resolve technical issues, protect voter privacy, and transparently resolve disputes, which promotes trust in the e-voting system.

No university representative or developer provided access to publicly unavailable parts of the source code of the i-voting application nor did an independent audit, therefore some of the results are declarations of interviewees without any possibility of verification. Given the only partial knowledge of the systems and the trade secrets of the developers, the rating is not definitive but is chosen based on all information gathered at the time of writing. Thus, due to the lack of transparency, compliance with **technical standards** is not sufficiently evaluated. Based on the given findings, final recommendations are made.

## 3.2. Questionnaire survey of university students and graduates focused on i-voting systems

In order to find out the opinions on i-voting of Czech students and graduates of universities, including academic staff, an online survey was conducted from 24 October 2023 to 12 May 2024 through the application *Dotaznik.czu.cz*, which uses the LimeSurvey system. Respondents answered questions defined in the Agbesi et al. (2023) framework focusing on Information Availability, Understandability, Monitoring and verifiability, Remedial Measures, Testing, Transparency, and several additional questions. All information, including consent for data processing, was displayed to participants on the survey homepage.

Respondents answered questions on a seven-point Likert scale ranging from Strongly disagree (0) to Strongly agree (6). A total of 177 people were recorded completing the questionnaire. A total of 108 questionnaires were completed in full. Of these, eight additional questionnaires were removed because the control question "This question is not part of the survey and just helps us to detect bots and automated scripts. To confirm that you are a human, please choose 'Strongly agree' here" other than Strongly agree. Of the 100 responses, 64 were male, 35 were female, and one respondent did not indicate gender. 72 respondents were aged 18-30, 18 were aged 31-40, five were aged 41-50, two were aged 51-60, and three were aged 61-70. 72 respondents have a high school degree, ten have a bachelor's degree, 12 have a master's degree, and six have a PhD.

The Statistica 14 software was used to produce the exact results of the descriptive statistics. Descriptive statistics were calculated separately for each variable, providing fundamental information such as the mean, minimum and maximum values, various measures of variation, and data regarding the shape of the variable's distribution (including the standard deviation and the standard error). An important aspect of the description of a variable was the shape of its distribution, which indicates the frequency of values within different ranges of the variables. More precise information was obtained by performing normality tests to determine the probability that the sample originated from a normally distributed population of observations, specifically using the Shapiro-Wilk test. These statistics were included in the dataset (Martínek & Tyrychtr, 2024).

For simplicity, we use in this paper the calculated average of the responses divided by the maximum (six), which allows us to present values on a scale from zero to one, where values above 0.5 to one (max) indicate the level of agreement with the question and values below 0.5 to zero (max) indicates the level of disagreement with the question. The value is rounded to three decimal places.

## 4. Results

Out of 26 Czech public universities, 18 institutions have already used the electronic system for remote voting in the Academic Senate elections in the past. In addition, three more, the Academy of Arts, Architecture and Design in Prague, Jan Evangelista Purkyně University in Ústí nad Labem, and the Institute of Technology and Business in České Budějovice, already have an electronic voting system in place, but it has not yet been used in the senate elections. Basic data on the address and developer of the systems used at these 18 institutions are described in the following table.

*Table 1: Systems for electronic election of the Academic Senate of Czech public universities (source: own elaboration based on interviews)*

| Public university | E-voting system (URL) | Developer |
|---|---|---|
| Academy of Performing Arts in Prague | volby.amu.cz | CVT UPOL |
| Brno University of Technology | vut.cz | VUT |
| Czech Technical University in Prague | volby.cvut.cz | ČVUT |
| Charles University | hlasovani.is.cuni.cz<br>volby.ff.cuni.cz | DERS Group<br>FA CUNI |
| Janáček Academy of Performing Arts | is.jamu.cz/auth/volby/ | FI MUNI |
| Masaryk University | is.muni.cz | FI MUNI |
| Mendel University in Brno | is.mendelu.cz/auth/evolby/ | IS4U |
| Palacky University Olomouc | volby.upol.cz | CVT UPOL |

| Prague University of Economics and Business | insis.vse.cz/auth/evolby/ | IS4U |
| Silesian University in Opava | is.slu.cz/auth/volby/ | FI MUNI |
| Technical University of Liberec | fraxinus.is.tul.cz/volby | TUL |
| Technical University of Ostrava | uzivatel.sso.vsb.cz/wps/myportal/uzivatel/volby-as/hlasovani/ | VŠB |
| Tomas Bata University in Zlín | volby.utb.cz | DERS Group |
| University of Chemistry and Technology, Prague | hlasovani.vscht.cz | DERS Group |
| University of Ostrava | portal.osu.cz/wps/portal/as | OSU |
| University of Pardubice | evolby.upce.cz | CVT UPOL |
| University of South Bohemia in České Budějovice | volby.jcu.cz | FA CUNI |
| University of West Bohemia | volby.zcu.cz | DERS Group |

For further evaluation, primarily Internet voting systems that have been choice tested at more than one Czech public university were considered. Specifically, these are the E-voting system from Faculty of Informatics Masaryk University in Brno (FI MUNI), the Elections system from Computer Center Palacky University Olomouc (CVT UPOL), the E-voting module of the UIS system from IS4U, the POLL system from DERS Group and the Elections system from Faculty of Arts Charles University in Prague (FA CUNI).

## 4.1. Regulations

In the Czech Republic, the election to the academic senates of universities is regulated by the Higher Education Act (Czech Republic, 2024), which defines that the election should be direct and secret. Direct suffrage means that members of the Senate cannot be elected by delegates. The procedure for election to the academic senate is addressed by universities and their faculties in the election rules of the academic senate or other internal regulations of the university. In the context of a remote electronic election in an uncontrolled environment, it is difficult to verify that the election was indeed direct, i.e. that the voter voted in person and did not, for example, hand over access data to another person. This issue is also addressed in the case of correspondence voting, where in both cases, even voting in an uncontrolled environment is generally accepted as direct in many cases. In order to verify that the election is indeed secret, at a minimum an independent security analysis of the system used should be carried out, and ideally the system should be open-source so that the method of secrecy can be transparently verified by the general public.

From the point of view of usability and accessibility, it is useful to distinguish between these two related but distinct concepts. Usability refers to how easily and efficiently the system can be used by all users. A highly usable system ensures that voters, regardless of their technical knowledge, can navigate the voting process smoothly and without confusion. Accessibility, on the other hand, specifically focuses on making the system accessible to people with disabilities, such as the visually impaired or those with physical limitations. An accessible system ensures that everyone, including people with disabilities, can fully participate in the voting process by providing them with necessary accommodations such as screen readers or keyboard navigation.

Also relevant in this context is the Law on Accessibility of Websites and Mobile Applications (Czech Republic, 2019), which obliges universities, as public institutions, to ensure the accessibility of content published on their websites. This Act also applies to content that a higher education institution is obliged to publish under other legislation, e.g. in the exercise of public administration in education, science, research, development, innovation and care for children and youth, or under the Act on Free Access to Information.

When it comes to the handling of personal data of constituents, universities must comply with the General Data Protection Regulation (EU, 2016). The GDPR applies directly in all EU Member States, in the Czech Republic it has been implemented through an amendment to some related laws. This ensures that the provisions of the GDPR are fully integrated into the Czech legal framework. Universities are obliged to handle personal data processed during the electoral process in accordance with these regulations, thus ensuring the protection and privacy of voters' data.

## 4.2. Democratic norms for i-voting

### 4.2.1. Participation

In the case of the student chamber of the Academic Senate, every person with student status has the right to vote, while every teaching staff member has the right to vote for the pedagogical chamber. As part of eligibility, universities use their own voter registry to verify users, where voter verification has already occurred when students register or begin employment. Each employee has 1 vote, regardless of the amount of working time. Any student or employee who registers as a candidate may be elected. A problematic situation arises when one person is both a student and a faculty member, where each college must determine the procedure so that a given voter can vote only once. Some online voting systems allow the voter to choose which chamber they want to vote for and then block the ability to vote for the other chamber of the Academic Senate. Elsewhere, a person is primarily placed in the academic staff chamber election unless they directly request to be placed in the student chamber. Voters' lists are normally approved by the designated Elections Committee before the election is run.

According to the number of full-time students, the largest faculty in the Czech Republic that has already used remote electronic voting for elections to the Academic Senate is the Faculty of Arts at Charles University in Prague, which has 5,560 full-time students (Ministry of Education of the Czech Republic, 2024). Electronic elections have been used at FA CUNI from 2016 to the present. The following graph (Figure 1) of voter turnout indicates that the implementation of remote electronic voting has not had a significant impact on voter turnout to define a clear trend. Of the years studied, the student chamber had the highest turnout of 12.7% in 1999 during the traditional paper ballot election, while the academic staff chamber had the highest turnout of 44% during the remote electronic election in 2019. Conversely, the lowest turnout during the study period for the student chamber was 3.2% in 2007 for the traditional paper ballot election and 15.8% for the academic staff chamber electronically in 2022 (Charles University, 2024). The higher student turnout at the turn of the millennium may be related to the minimum quota for validity in place at the time, with faculties attempting to increase student participation in various ways to ensure the validity of elections.

*Figure 1: Chart of voter turnout at FA CUNI from 1999-2022, remote electronic voting introduced in 2016 (source: own processing according to Charles University (2024))*

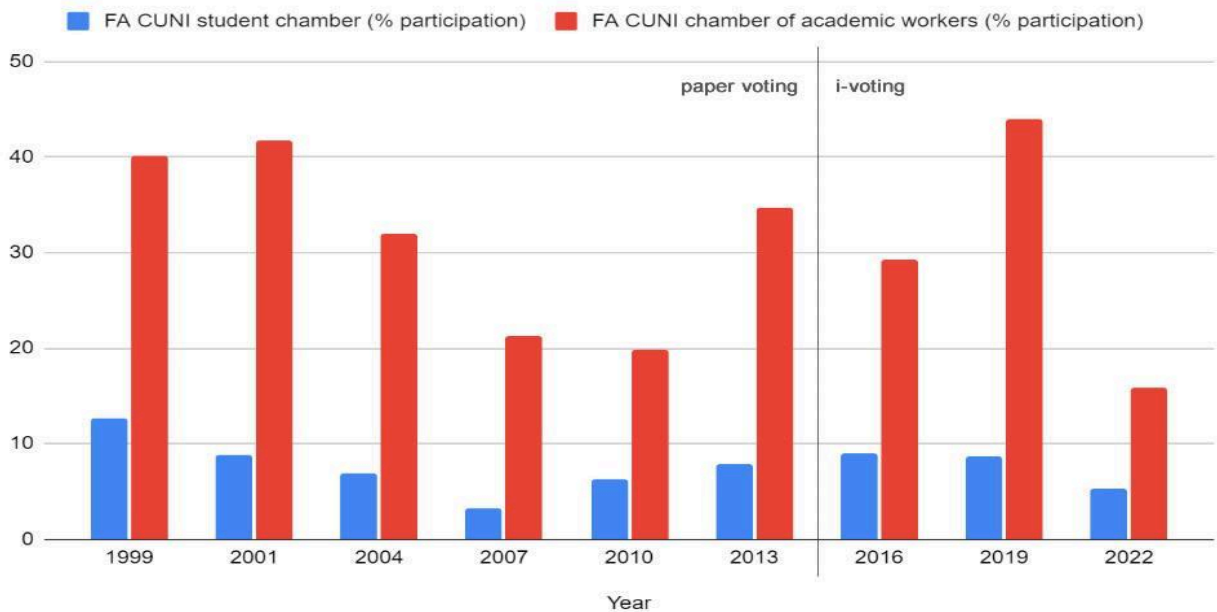

In the survey, the agreement rate was 0.723 for the question that respondents would use remote electronic voting if this option were available in the event of an election.

### 4.2.2.   Representation

Within the academic senates, the ratio of representatives of the teaching chamber to representatives of the student chamber usually prevails; at Charles University, the ratio of academic staff and students is balanced. In the case of elections to the university academic senate, inequality arises because each faculty usually has a certain number of representatives, which may not sufficiently reflect the size of the faculty. In faculty academic senate elections, each student usually has an equal vote for student representation, and each teacher has an equal vote for student representation. However, some faculty academic senates also have a split number of representatives for each discipline or department under the faculty. The CVT system of UPOL allows combining the election for both chambers, so it is possible to vote for up to 6 academic staff and one student in the pedagogical curia, while in the student curia, the voter can vote for up to two academic staff and three students. The possibility of selecting candidates and the method of determining the order in the CVT UPOL i-voting system is shown in Figure 2.

*Figure 2: Candidate selection in the CVT UPOL i-voting system, 1 is the list of candidates with the option to remove, 2 indicates a change in order, 3 indicates proceeding to submit the ballot (UPOL, 2022)*

Most institutions allow any nominee who agrees to run to run. Mendel University in Brno also conducts a remote electronic nomination round where each member of the academic community can nominate up to ten candidates for the Academic Staff Chamber or up to five candidates for the Student Chamber. After the nomination round is completed, the results are published and requests for consent to candidacy are sent out to the nominees. Subsequently, lists of candidates are drawn up on the basis of the nomination support, with no more than twice as many candidates who have consented to stand as the number of Academic Senate seats to be filled. Only then is there an election round.

### 4.2.3.  Competitiveness

Candidates are nominated by members of the academic community. Each candidate who agrees to be nominated normally submits basic information about themselves and their election program, which is normally announced on the institution's website. An example of filling in the ballot paper, which contains not only the name of the candidates but also their photo, job position and a link to more detailed information, in the FI MUNI i-voting system is shown in Figure 3.

*Figure 3: Candidate selection in the FI MUNI i-voting system, where it is also possible to view the candidates' photos, the option "I abstain from voting" is displayed and the arrow indicates the button to save the vote (SLU, 2020)*

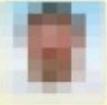

### 4.2.4. Transparency (and accountability)

The internet voting systems used at Czech universities are not very transparent, the source codes of the voting systems used are not published to the public, and voters usually only have access to basic documentation containing mainly voting instructions.

Responsibility for the election rests with the election committee established by the Academic Senate. Individual faculties of some universities usually have to adapt to the election method determined by the university academic senate, so the university academic senate may also have some responsibility for choosing the election method.

In the questionnaire survey, for the question that the online voting system should be transparent, the level of agreement of respondents is 0.807. Transparency as an integral part of the Internet voting system is reflected in an agreement rate of 0.778. Transparency as an important feature of the internet voting system is expressed by an agreement rate of 0.785.

### 4.2.5. Responsiveness

Elections at Czech universities are secret, so the elected senator has no clear idea which voters' preferences he or she represents. Candidates may have election programs that are often brief or missing altogether. All voters have the option of directly approaching academic senators to whom they can convey their demands, who then represent them as they see fit. Meetings of the Academic Senate are normally open to the public and records of proceedings are openly published. The activities of each senator are then reflected by voters in the next election.

The election results are generally accepted once the protocol has been certified by the Electoral Commission. As a rule, voters and candidates have the opportunity to lodge a complaint about the conduct of the election. Once the complaints have been settled by the Electoral Commission, the results are considered valid and the newly elected Senators take over the certificates of election and take office from the beginning of the new term of the Academic Senate.

The introduction of remote electronic voting has also encountered problems at universities in the Czech Republic. After the remote electronic elections to the Academic Senate of Charles University in Prague at the Faculty of Science in 2006, a complaint was filed that the elections were invalid due to their form not complying with the existing regulations, which was upheld by the main election commission by invalidating the elections (AS PřF UK, 2007).

In the survey, respondents report an agreement rate of 0.735 on the question that a responsible party should be identified in case of any problems with the electoral process.

## 4.3. Technical standards for i-voting

### 4.3.1. Security

Because the voting systems used in Czech public universities are not transparent, in that they do not publish source code or detailed information about their operation, it is difficult to verify their security and integrity. Based on the interviews conducted, only basic information about the functioning of the systems is available for research, which cannot be independently verified. The systems in use eliminate the possibility of inadvertently filling in a ballot paper by mistake; some leave the option of submitting a blank ballot paper. All systems examined use Hypertext Transfer Protocol Secure (HTTPS) for secure connections.

The Prague University of Economics and Business states that it has had an independent security audit of the system carried out according to the methodology of the National Office for Cyber and Information Security (NÚKIB, 2023). The audit, conducted by a company certified to ISO security standards, also included vulnerability tests from both internal and external networks. The audit did not reveal vulnerabilities, but it is not publicly available.

The CVT UPOL's Voting system uses asymmetric encryption for sending the vote, where the vote is encrypted already in the voter's device, the votes cannot be manipulated externally - the database antitampering is at the level of symmetric encryption and checksums over the database, so it is possible to detect possible violations of election integrity in the database or in the application. The application symmetrically encrypts the votes on the database and continuously generates checksum. Check is done when the election is evaluated or can be done as needed. Organizationally, this is ensured by the fact that the database and web server administrator does not know the symmetric key that is stored in the application; the developer, on the other hand, does not have access to the web server and database.

*Figure 4: Example of DERS Group's POLL electronic ballot with the Abstain option, the arrow indicates the recap button where the choice must be confirmed before submitting (ZČU, 2023)*

In the event of an outage, systems including POLL (DERS Group, 2021) allow for extended voting at the discretion of the Election Commission. All operations including importing lists, modifying lists, and exporting results are logged within the POLL system for possible review. An example of an abstention vote and requiring confirmation in the POLL i-voting system is shown in Figure 4.

An agreement rate of 0.85 respondents presents the requirement for there to be an assurance that any security breach will be detected and repaired. The requirement for thorough testing for security vulnerabilities by independent security experts is represented by a high agreement rate of 0.89.

### 4.3.2. Verification process and secrecy

Single sign-on credentials are used to log in to the electoral system and are also used by academic staff and students to access other College systems. The system from DERS Group uses Eduid.cz operated by CESNET, where all Czech public universities are full members of the association. The IS4U system allows setting up multi-factor authentication, For example, Prague University of Economics and Business requires two-factor authentication from 2023. Specifically, after entering the correct login name and password, the user is prompted to enter an authentication code from the mobile app. If the user designates the device as secure, the user is not required to provide a login code for 30 days.

In order to increase coercion resistance, the FI MUNI system offers the possibility of repeated voting, where only the last vote cast is counted. However, most of the systems used at Czech universities do not allow repeated voting, so the only way to reduce the risk of pressure from colleagues and others is to cast your vote early.

*Figure 5: The page after selecting the options in the IS4U system, where the voter is informed that the ballot has been encrypted in the voter's browser, the ballot's digital signature string is displayed to check the vote count, and the button to cast the encrypted vote is displayed (VSE, 2021)*

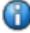

*Figure 6: Post-submission page showing the vote control code in the UPOL CVT system (UPOL, 2022)*

The systems from FI MUNI, CVT UPOL (Figure 6), or IS4U (Figure 5; Figure 7) allow to individually verify the counting of the sent vote. Universal verifiability for independent verification of election results including their integrity and ensuring high credibility of elections is not offered by standard systems at Czech public universities.

*Figure 7: IS4U ballot box page with a list of digital signatures of individual votes with the option to switch to the voters list (VSE, 2021)*

Due to the lack of transparency, we were unable to determine the method of vote secrecy for most of the systems used to ensure voter privacy. For example, the POLL system is based on documents mandatorily published in the Register of Contracts (DERS Group, 2021), where, according to the technical requirements of the contract, confidentiality is addressed after submission by storing the anonymized vote and indicating that the voter has already voted. The voter subsequently receives information that the vote was saved.

Within the survey, a high agreement rate of 0.885 is recorded for the question asking for the ability to verify that the online voting system recorded the voter's vote as intended. The requirement for public availability of external audits after the election is represented by an agreement rate of 0.79.

### 4.3.3. Data Ownership and Control

Some Czech higher education institutions such as FA CUNI or FI MUNI develop voting systems for their school and possibly other universities on their own. In this case, they have full control over the data and the whole system. On the other hand, universities that use the services of IS4U or DERS Group rely on a commercial system operator by way of a contractual relationship. In most cases, the systems run on servers located in the Czech Republic. In terms of election administration, it is not common for the administration to be distributed - in many cases, one person sets up the elections in the system. FI MUNI's I-voting system allows the generation of a final report for each election, as shown in Figure 8.

*Figure 8: Election result log generated by the FI MUNI E-voting system, the log includes the name, description, and date of the election, as well as the name of the election administrator, importance, results, and other information (MUNI, 2024)*

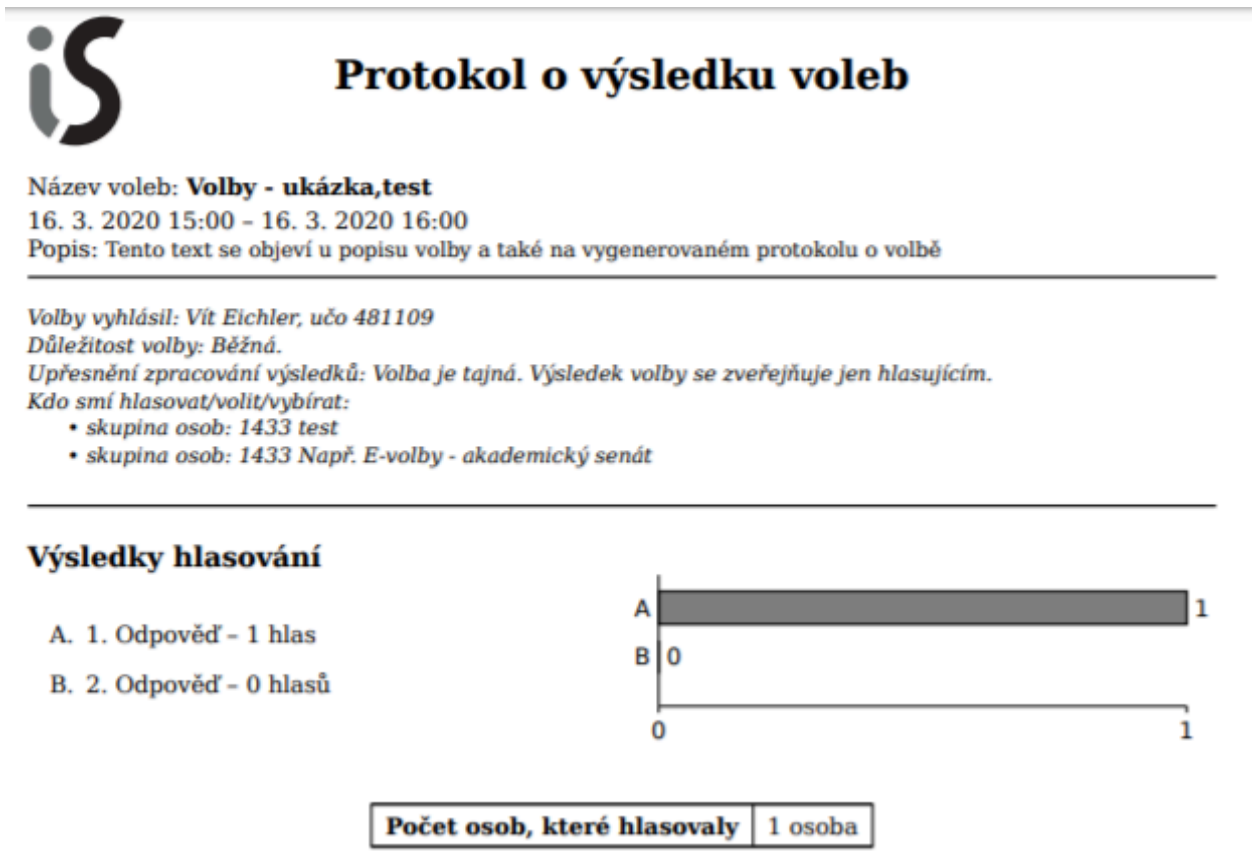

## 4.4. Overall assessment of compliance with democratic principles in the use of i-voting systems at Czech public universities

From the available information, it cannot be clearly assessed whether the i-voting systems used in Czech public universities meet basic democratic requirements. Universities and developers do not publish the source codes of the applications used, offer only weak documentation, and are often unwilling to disclose the information needed to verify how the i-voting systems work. This information is not only unavailable to the general public, but also to the eligible voters themselves, who must therefore blindly trust i-voting systems. The interviews show that often even the responsible representatives of the universities do not know more detailed information about the functioning of the i-voting systems, and thus again put their full trust in the developers and operators. The current legislative regulation does not force universities or developers to publish substantial information about the functioning of i-voting systems, even though the specifics of the electoral process require such public information in a democratic society.

Based on a basic descriptive evaluation and the assumptions required by potential voters, a system should be used that ensures security, privacy, verifiability, voting integrity, and transparency. Since no i-voting system used in Czech universities is fully transparent, a publicly accessible independent audit of the system would have to be made to guarantee these properties. Systems with individual verifiability from FI MUNI, CVT UPOL or IS4U are close to meeting the expected requirements.

## 5. Discussion

In contrast to the transparent approach of Adida et al. (2009), it is difficult to access information in Czech public universities. The documentation is very limited, the persons in charge only answer basic questions in general or not at all, and it is often not clear whether the way i-voting systems work is at least known by the persons in charge at the universities. Without adherence to democratic norms involving transparency, the i-voting systems used in Czech universities cannot be sufficiently evaluated in terms of compliance with technical standards involving privacy, security or election integrity. On the other hand, a detailed evaluation can be carried out for the system of the Czech Pirate Party, which publishes all source code and other necessary information about its i-voting system (Martínek & Malý, 2024). I-voting systems differ to a large extent from other information systems, where a possible error or manipulation can be discovered and corrected. Prioritizing electoral efficiency over electoral transparency in the case of a "black box" i-voting system is not desirable (Cheeseman et al., 2018), as suggested by potential users of i-voting systems themselves in our survey. Interviewees often justified their reluctance to provide sufficient information by citing system security. In some cases, they themselves did not have information on how the system in question ensures its functionality. The lack of transparency needed to evaluate the system has a negative impact on the credibility of the electoral process.

The security analyses of the Helios system (Bernhard et al., 2012; Cortier & Smyth, 2013; Acemyan et al., 2015; Panizo Alonso et al., 2018) used at the University of Leuven allow correcting the errors found and building confidence in the integrity of the election among its users. There are no publicly available security analyses regarding the software used in Czech public universities. It is worth adding that similarly low transparency exists in other universities around the world. For example, Harvard University offers more detailed information regarding remote electronic voting (Harvard University, 2024), including a link to the operator of the voting system (ESC, 2024), which offers a basic FAQ and other materials, but the source code of the system or independent audits are not publicly available.

Despite the lack of transparency of i-voting systems at Czech public universities, the practice of remote voting via the Internet for the Academic Senate is widespread, with 18 out of 26 Czech public universities using i-voting systems. These platforms can serve the purpose of reducing costs and administrative burdens. However, it is quite concerning given that these platforms are being used to elect individuals to significant positions, which manage large financial budgets and oversee faculties or universities. At the same time, even potential voters themselves in the questionnaire survey express a high level of agreement with the requirements for transparency, verifiability, clarity or security.

Similar to studies in national elections (Germann & Serdült, 2017; Goodman & Stokes, 2020), it is difficult to determine whether the introduction of i-voting at universities affects voter turnout; we do not find a clear trend in the university turnout we observe.

In order to enable the general public to scrutinize their i-voting systems and to ensure the credibility of democratic elections, the possibility of developing a recommendation, following the example of the French CNIL (La Commission nationale de l'informatique et des libertés, 2019), which would require public institutions, including public schools, to comply with the basic requirements for a secret online voting system where the outcome of the vote has a significant impact, such as in the case of the election of members of the Academic Senate. A national recommendation with binding requirements for public administrations could be based on Recommendation CM/Rec(2017)5 of the Committee of Ministers of Member States on standards for electronic voting (Council of Europe, 2017), which is a valid recommendation for Council of Europe member states.

Inspiration can also be found in the transparently managed introduction and control of i-voting systems in the context of politically binding elections. In the implementation of a remote e-voting project in Norway, a transparent tender was conducted (Ministry of Local Government and Regional Development, 2010), where an additional open source requirement allowed for a technical analysis of the system (Bjørstad, 2013), which discovered vulnerabilities for remediation. Similarly, in Estonia, most of the source code of the system in use is made public (Valimised, 2024). Transparency allowed security analyses to be conducted (Springall et al., 2014; Heiberg et al., 2015; Pereira, 2023), based on which modifications could be made to the system to improve its security. Thus, in Estonia, the system is regularly updated to ensure greater security and integrity of the elections, which is intended to increase the credibility of the entire remote e-voting system. A specific description of the possibility of implementing a transparent open-source i-voting solution including its verification is described in Martínek et al. (2024).

The recommendation for academic senates and university administrations is that they require open-source systems, or at least have the possibility of an independent audit of the system, including access to the source code by experts who can sign a non-disclosure agreement (NDA). In the event that student and academic staff constituents do not have sufficient assurance that the i-voting system is functioning properly, they should demand that elected representatives to the Academic Senate provide information or an audit. A recommendation for the state may be to require public institutions within i-voting systems to comply with Recommendation CM/Rec(2017)5 of the Committee of Ministers of Member States on standards for electronic voting (Council of Europe, 2017). Nevertheless, it should be borne in mind that even Recommendation CM/Rec(2017)5 of the Council of Europe is not the final perfect list of requirements and needs updating (Rodríguez-Pérez, 2022).

# 6. Conclusion

This paper has provided a comprehensive analysis of the implementation of i-voting systems at Czech public universities, applying the evaluation framework proposed by Von Nostitz et al. (2021). The findings highlight several key areas for improvement, particularly in terms of transparency and security. While the adoption of i-voting systems has undoubtedly improved convenience and accessibility for voters, these systems often lack sufficient transparency, which is crucial for maintaining voter confidence in the integrity of the electoral process.

The analysis also underscores the need for universities to ensure that their i-voting systems meet both democratic and technical standards, particularly as these systems are used to elect representatives to significant positions within university governance. Recommendations for enhancing the transparency of these systems include the adoption of open-source platforms or, at a minimum, allowing independent audits of the systems used.

Overall, while i-voting systems have the potential to streamline the electoral process, the findings indicate that without improved transparency and security measures, these systems may undermine the democratic legitimacy of elections at Czech universities. Future research should focus on conducting more in-depth evaluations of these systems and exploring ways to further improve their adherence to democratic principles.